\documentclass[letter ]{aa}  

\usepackage{graphicx}
\usepackage{upgreek}
\usepackage{txfonts}
\usepackage{adjustbox}
\usepackage{color}
\usepackage{amsmath}
\usepackage{amssymb}
\usepackage{breqn}
\usepackage{gensymb}
\usepackage{placeins}
\usepackage{natbib}
\usepackage{subfig}
\usepackage{multirow}
\usepackage{booktabs}
\usepackage{colortbl}
\usepackage{csvsimple}
\usepackage[colorlinks=true,
linkcolor=blue,
citecolor=magenta]{hyperref}
\usepackage{txfonts}
\usepackage{comment}
\usepackage{gensymb}
\usepackage{pgfplots}
\newlength\figureheight
\newlength\figurewidth
\usepackage{orcidlink}
\usepackage{soul}

\usepackage{calligra}
\usepackage[T1]{fontenc}

\begin{document}

\title{The infrared jet of M87 observed with JWST}

\titlerunning{The infrared jet of M87 observed with JWST}

\author{%
Jan R\"oder\inst{\ref{iaa}}\orcidlink{0000-0002-2426-927X} \and 
Maciek Wielgus\inst{\ref{iaa}}\orcidlink{0000-0002-8635-4242}\and
Joseph B. Jensen\inst{\ref{jensen}}\orcidlink{0000-0001-8762-8906}\and
Gagandeep S. Anand\inst{\ref{anand}}\orcidlink{0000-0002-5259-2314}\and
R. Brent Tully\inst{\ref{tully}}\orcidlink{0000-0002-9291-1981}
}

\authorrunning{J. R\"oder et al.}

\institute{
    Instituto de Astrofísica de Andalucía, Gta. de la Astronomía, 18008 Granada, Spain\\\email{contact@janroeder.science}\label{iaa}
    \and
    Department of Physics, Utah Valley University, 800 West University Parkway, Orem, UT 84058, USA \label{jensen} 
    \and
    Space Telescope Science Institute, 3700 San Martin Drive, Baltimore, MD 21218, USA\label{anand}
    \and
    Institute for Astronomy, University of Hawaii, 2680 Woodlawn Drive, Honolulu, HI 96822, Hawai'i \label{tully}
}

\date{Received 24 July 2025/ Accepted XX YY ZZZ}

\abstract
{}
{We present the first JWST+NIRCam images of the giant elliptical active galaxy M87 and its jet at 0.90, 1.50, 2.77 and 3.56\,$\upmu$m. We analysed the large-scale jet structure, identifying prominent components, and determined the near-infrared spectral index.}
{The data were calibrated using the standard JWST pipeline. We subtracted a constant background level and a smooth model of the galaxy surface brightness to isolate the jet.}
{The total image fluxes measured in the NIRCam filters follow the infrared bump pattern seen near 1.6\,$\upmu$m in the spectral energy distribution of M87, caused by the surrounding stellar population in the galaxy. The residual jet images broadly agree with the radio to optical synchrotron power law $S_\lambda \propto \lambda^\alpha$ with $\alpha= 0.7$--1.0. We identified the most upstream knot L at a distance of ($320\pm 50$)\,mas from the core. The component HST-1, at ($950\pm 50$)\,mas from the core, is transversely resolved, and both the individual images and the spectral index map clearly indicate its double-component substructure with two elements of similar size and flux density, with centroids separated by ($150\pm 20$)\,mas and with a significantly larger spectral index $\alpha$ observed in the downstream component ($\alpha_{\rm do} = 0.30$) than in the upstream one ($\alpha_{\rm up} = -0.15$). We also observe the counter-jet component located about 24\,arcsec away from the nucleus.}
{}


\maketitle

\section{Introduction}

The giant elliptical galaxy M87 \citep{Messier1781} has a long history as one of the most intensely studied objects in the sky \citep[for a review see, e.g.,][]{Hada2024A&ARv..32....5H}. The prominent one-sided jet, discovered just over a century ago \citep{Curtis1918}, was eventually found to be polarized \citep{Baade1956ApJ...123..550B}--leading to the initial conclusion that synchrotron radiation had to be the dominant emission mechanism at optical wavelengths.

The M87 jet has since been a prime laboratory for the science of active galactic nuclei (AGN). Since M87 is a strong radio source, many studies of its jet at various spatial scales were carried out at radio wavelengths \citep[e.g.,][]{Baldwin1956Obs....76..141B,Reid1989ApJ...336..112R,Lobanov2003NewAR..47..629L,Kovalev2007ApJ...668L..27K,Chang2010A&A...515A..38C,Mertens2016A&A...595A..54M,Britzen2017A&A...601A..52B,Kim2018A&A...616A.188K,Punsly2023A&A...677A.180P,Cui2023Natur.621..711C,Nikonov2023MNRAS.526.5949N}, even going as far as resolving the central supermassive black hole at its core \citep{M87p1,M87_2018_p1,Lu2023,Kim2025A&A...696A.169K}. 
Many other studies examined the optical emission \citep[e.g.,][]{Boksenberg1992A&A...261..393B,Perlman1999AJ....117.2185P,Perlman2001ApJ...551..206P,Perlman2003ApJ...599L..65P,Meyer2013ApJ...774L..21M}, UV radiation \citep{Madrid2007Ap&SS.311..329M,Lessing2024ApJ...973..144L}, X-rays \citep[e.g.,][]{Schreier1982ApJ...261...42S,Marshall2002,Snios2019ApJ...879....8S} and $\gamma$-rays \citep{HESS2024A&A...685A..96H}. 
The multi-band emission was studied by \citet{Harris2009ApJ...699..305H}, \citet{Prieto2016}, and \citet{EHTMWL2021,EHTMWL2024}.

Previous optical and infrared studies found that the radio-to-optical spectrum in the kpc-scale jet is generally well described by a non-thermal synchrotron power law \citep{Smith1983ApJ...266...69S,Perlman2007ApJ...663..808P}. In this work, we present the first JWST+NIRCam infrared images of the kpc scale jet of M87. We adopted a distance estimate $D_{\rm M87}=16.3$\,Mpc from \citet{Anand2025}, 
derived from the same JWST data set, consistent with \citet{Bird2010A&A...524A..71B} and \citet{M87p1},
such that 1\,arcsec\,$\approx$\,80\,pc projected distance. 

\section{Observations and data reduction}
\label{sec:obs+data}

\begin{figure}
    \centering
   \includegraphics[trim={10cm 13cm 13cm 10cm},clip,width=0.48\columnwidth]{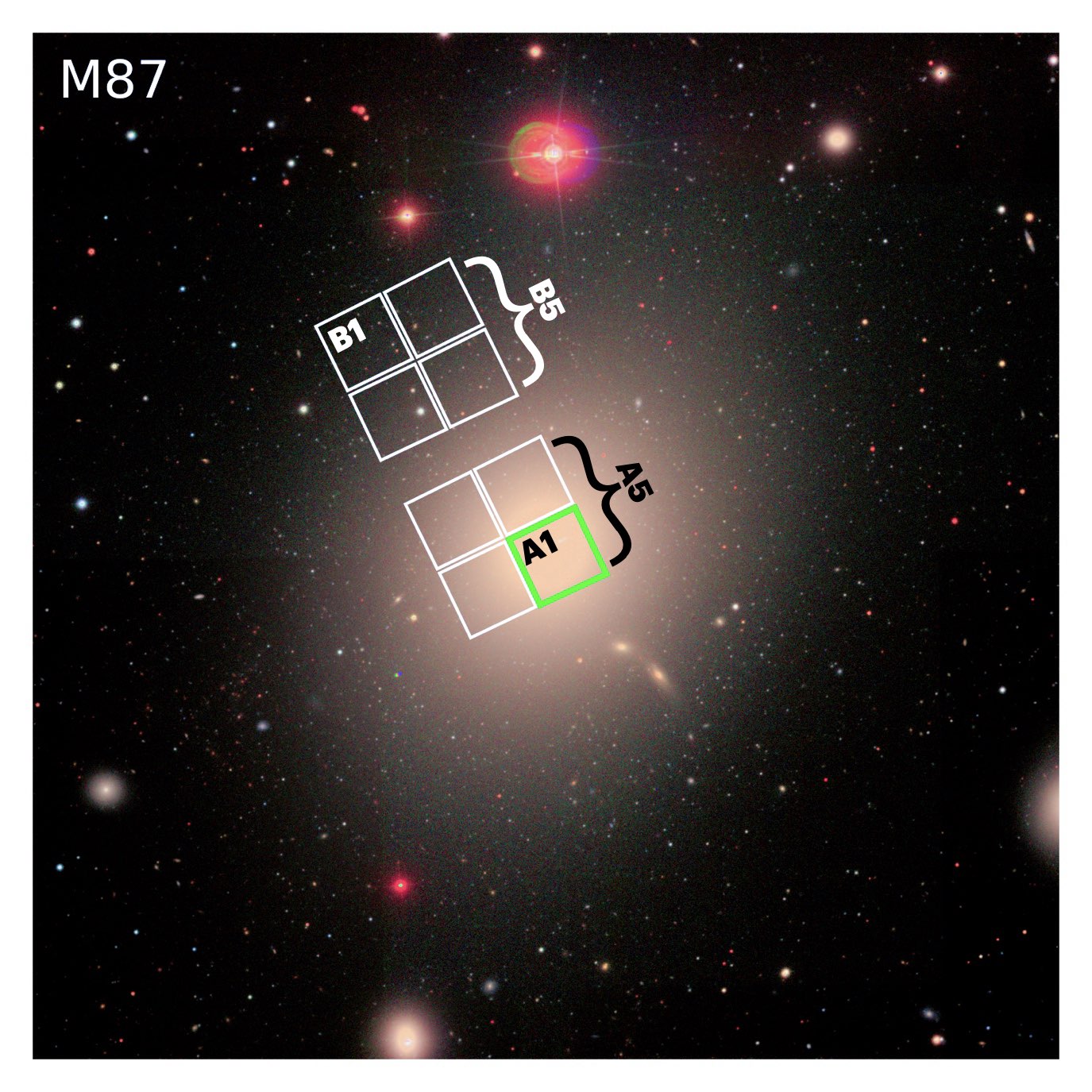}
     \includegraphics[width=0.48\columnwidth]{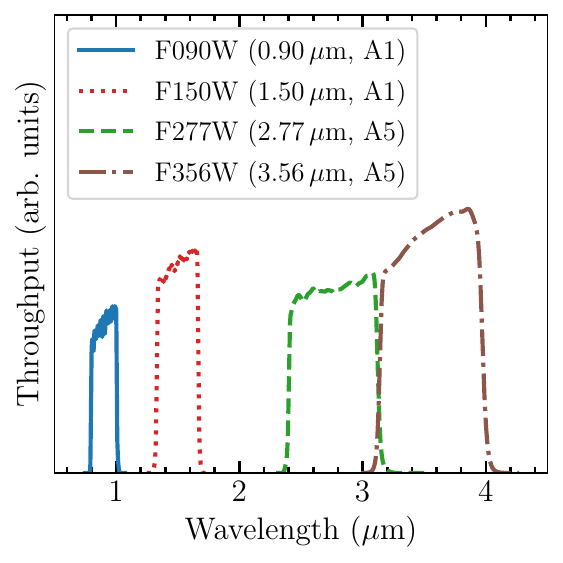} 
    \caption{\textit{Left:} NIRCam footprint on M87 \citep{Anand2025}. The green sub-square represents the A1 detector footprint for the F090W and F150W images in this study; F277W and F356W images are taken with the lower-right half of the NIRCam footprint (detector A5). The background image was created using the online SDSS mosaic tool. \textit{Right:} Detector throughputs and wavelength ranges for the four NIRCam filters used. Characteristic wavelengths are indicated in the legend.}
    \label{fig:NIRCam}
\end{figure}

M87 was observed with JWST+NIRCam in June 2024 as part of a program containing 13 other elliptical galaxies, to calibrate surface brightness fluctuations (SBF) using the tip of the red giant branch (TRGB) stars.\footnote{\url{http://dx.doi.org/10.17909/6rss-ck61}} The NIRCam footprint on the galaxy and the detector bandwidths are shown in Fig. \ref{fig:NIRCam}. The 64\,arcsec\,$\times$\,64\,arcsec A1 detector (0.031\,arcsec/px) was exposed for 4\,$\times$\,1460.2\,s and 4\,$\times$\,311.4\,s using the F090W and F150W filters, respectively. The same respective exposure times were used for F277W and F356W on the A5 detector (129\,arcsec\,$\times$\,129\,arcsec, 0.063\,arcsec/px).

The raw \mbox{NIRCam} images were processed using the standard JWST pipeline \citep{Bushouse2020ASPC..527..583B} to the 2b {\tt _i2d.fits} level, including instrumental corrections such as flat fielding, flux (ramp) photometric calibration, bad and saturated pixel correction, and image distortion correction using bilinear (flux-conserving) interpolation. The images for each wavelength were created by stacking four dithered exposures, each offset by a few pixels to allow for correction of bad pixels and cosmic rays; the offsets were, however, not large enough to correct for the large patch of bad pixels in the A1 detector located inconveniently near the end of the jet. 

The exposure times were chosen to achieve the depth needed to detect the red giant branch stars in the outer fields and to minimise the data volume; as a consequence, the pixels covering the galaxy centre are saturated. The combined images were scaled\footnote{\url{https://jwst-docs.stsci.edu/jwst-near-infrared-camera/nircam-performance/nircam-absolute-flux-calibration-and-zeropoints}} to units of Jy/arcsec$^2$; the uncertainties in the photometric zero points are currently estimated to be 1\%--2\% in individual detectors at the shorter wavelengths and 2\%--4\% in the longer wavelength filters.\footnote{\url{https://jwst-docs.stsci.edu/jwst-calibration-status/nircam-calibration-status/nircam-imaging-calibration-status}}

To isolate the jet, the dominant galactic emission had to be removed (see Appendix~\ref{sec:processing}). We first subtracted a constant background level for each filter, measured in the most distant detectors from the galaxy centre (B1 and B5, respectively, which correspond to the upper-left corner and upper left half of the footprint in Fig.~\ref{fig:NIRCam}). 
Then, we masked extraneous objects, such as foreground stars, dust in the galaxy, background galaxies, globular clusters, and the jet before modelling the galaxy at each wavelength using the \texttt{elliprof} program \citep{Tonry_1997, Jordan_2004}. \texttt{elliprof} fits the galaxy surface brightness in concentric elliptical annuli moving out from the centre, past the end of the jet, and extrapolates the model to the edges of the frame.
Lastly, we  fitted a smooth residual background function on a 15\,$\times$\,15 pixel grid to ensure that the global background level was exactly zero, and that smooth residual patterns from the galaxy subtraction process were removed.

\section{Results}

\subsection{Jet structure}

\begin{figure}
    \centering
    \includegraphics[width=0.98\columnwidth]{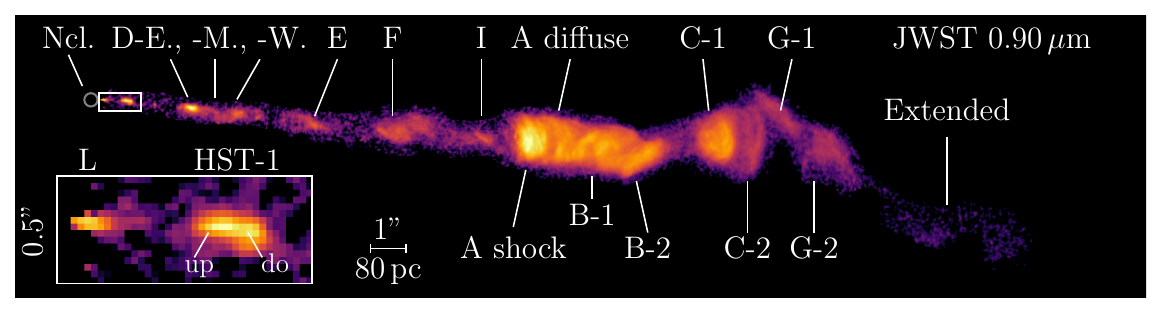}
    \includegraphics[width=0.98\columnwidth]{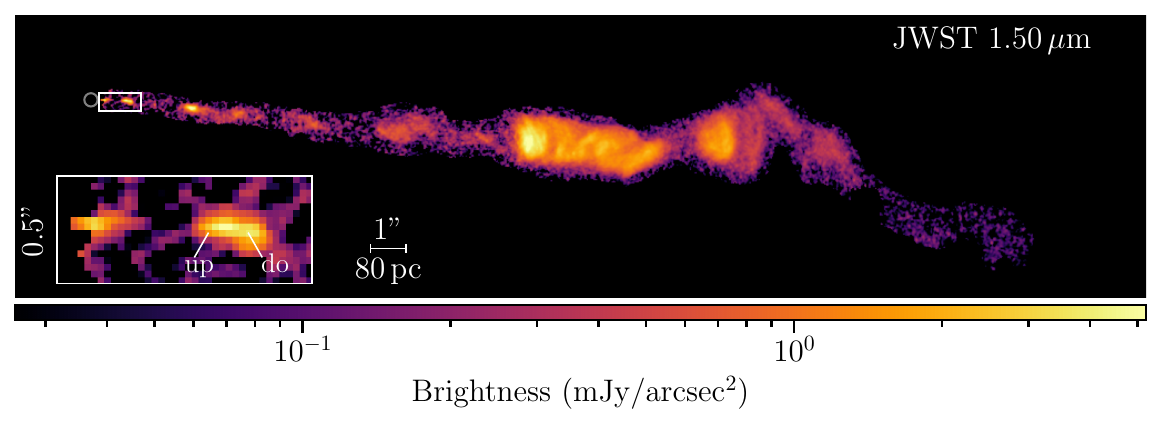}
    \includegraphics[width=0.98\columnwidth]{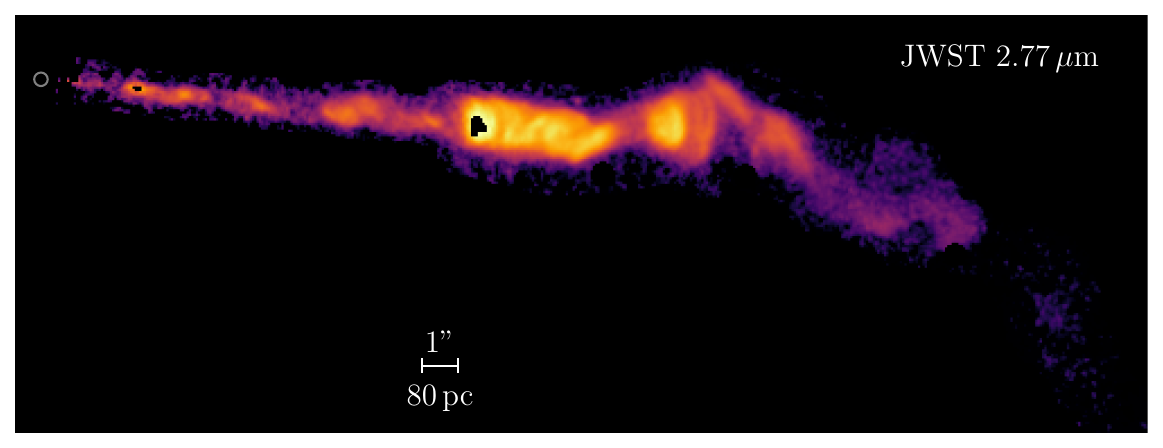}
    \includegraphics[width=0.98\columnwidth]{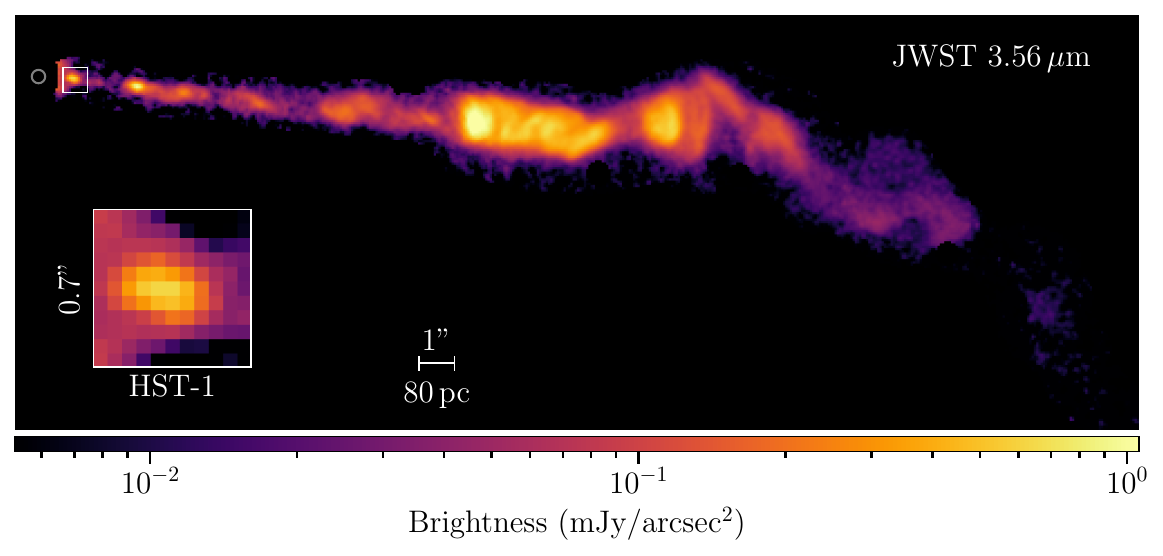}
    \caption{Masked residual jet images with dynamic range of 200 and limits chosen to maximize visibility of individual features. The first component in the longer-wavelength images (185\,$\times$\,502~pixels) is \mbox{HST-1}. The shorter-wavelength images (255\,$\times$\,1020~pixels) contain the most upstream component L.  }
    \label{fig:jet}
\end{figure}

Figure \ref{fig:jet} shows the galaxy-subtracted and masked residual jet images for the four wavelength bands. There are saturated pixels in the nucleus and, in the case of 2.77\,$\upmu$m, in some bright jet knots. It was therefore not possible to extract information from these regions. 
Before masking, the image background appears very bumpy (see Appendix \ref{sec:processing}); this is not detector or background noise, but rather fluctuations in surface brightness caused by the Poisson statistics of the discrete number of red giant stars per pixel. On small scales of a few pixels, the surface brightness fluctuations contribute to flux randomly and only average out on scales significantly larger than the point spread function.

We identify the individual jet components adopting the naming convention from \cite{Biretta1999} and \citet{Perlman2001ApJ...551..206P}. 
The centroid of the innermost distinct component, which we associate with the slowly moving knot L of \citet{Biretta1999}, is located at ($320\pm 50$)\,mas downstream from the core. We estimated the component size of 60\,mas\,$\times$\,150\,mas (full width at half maximum of a fitted elliptical Gaussian), elongated in the direction of the jet, with a flux density of ($25\pm 10$)\,$\upmu$Jy measured at 1.50\,$\upmu$m.

At the estimated distance of ($950\pm 50$)\,mas from the nucleus sits HST-1, the fastest-moving superluminal feature on these scales \citep[up to 24.1\,mas/yr;][see also \citealt{Chang2010A&A...515A..38C,Thimmapa2024}]{Snios2019ApJ...879....8S}. HST-1 is transversely resolved, with a fitted size of 80\,mas\,$\times$\,230\,mas, and is tilted about 10$^\circ$ away from the jet axis.  
A substructure of two similarly sized components is resolved well, with ($150\pm 20$)\,mas separation between the brightness centroids and the downstream component shifted off the jet axis. 
The total measured flux density of HST-1 is ($110\pm 20$)\,$\upmu$Jy and the inner component is brighter by around 30\% compared to the outer one at 1.50\,$\upmu$m. The sub-components blend together at longer wavelengths, see the last panel of Fig.~\ref{fig:jet}.

HST-1 is followed by component D, split into its Eastern, Middle, and Western parts. D-East is the brightest, and is connected to D-West by a "twist" denoted as D-Middle. The individual parts of D, as well as the more diffuse component E, move at mild superluminal speeds \citep{Biretta1995ApJ...447..582B,Biretta1999}. Component F has a clear double structure in the two shorter-wavelength images, and component I appears distinct, but both are blended into the jet emission at the longer wavelengths. The compound of components A and B features the bright shock \citep{Perlman2001ApJ...551..206P} followed by more diffuse emission, as well as a subsequent bend composed of components C and G. The jet fades out into extended emission, more apparent in F277W and F356W. 

Apart from the prominent forward jet, we detect signatures of a decelerating counter-jet around 24\,arcsec from the nucleus \citep{Sparks1992Natur.355..804S, Stiavelli1992Natur.355..802S}. In the 2.77 and 3.56\,$\upmu$m images, the counter-jet consists of two filaments connected by a hotspot, forming a C-shape, consistent with its morphology in the radio wavelengths \citep{Pasetto2021ApJ...923L...5P}. In the 0.90 and 1.50\,$\upmu$m images, only a faint trace of the hotspot is detected, and the filaments are not visible (see Appendix~\ref{sec:counterjet} for details).

\subsection{Near-infrared spectral index}\label{sec:NIRspec}

\begin{figure}[!htp]
    \centering
    \includegraphics[width=\columnwidth]{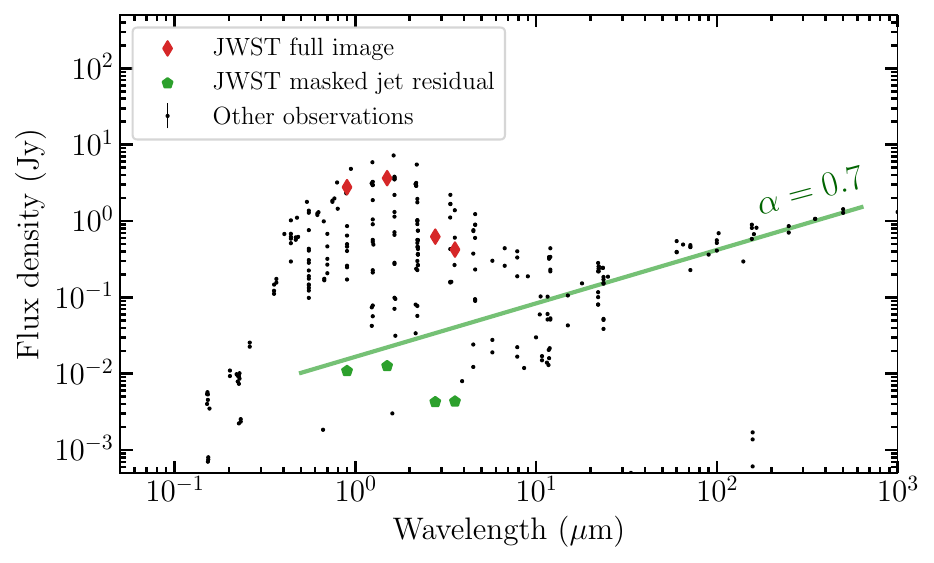}
    \caption{Optical and infrared spectrum of M87 with the total and residual image fluxes obtained from the JWST observations overplotted. The spectrum was created with SSDC building tools using data from the NASA/IPAC Extragalactic Database.}
    \label{fig:SED}
\end{figure}

Figure \ref{fig:SED} shows the optical and infrared spectrum of M87, centred around the 0.1--10\,$\upmu$m infrared bump caused by the \textsc{H}$^-$ opacity minimum in cool stars \citep{Sorba2010ApJ...721.1056S}, peaking at $\sim$\,1.6\,$\upmu$m. Our shorter-wavelength observations fall before the peak of the bump; the flux density then falls sharply between 1.50\,$\upmu$m and 2.77\,$\upmu$m across the peak, following the spectrum of the cool, old, low-mass stars in the giant elliptical galaxy \citep{BC-1993ApJ...405..538B}.

\begin{table}
\caption{Image-integrated flux densities}              
\label{tab:fluxes}     
\centering                                    
\begin{tabular}{ccccc }         
\hline\hline                      
Filter & Total & Model & Background  & Jet (masked)\\  
  & Jy & Jy &  10$^{-9}$\,Jy/px  & mJy\\  
\hline                                  
    F090W  & 2.78 & 2.74 & 31.10   & 10.89 \\    
    F150W  & 3.65 & 3.62 & 26.10   & 12.62 \\    
    F277W  & 0.63 & 0.62 & 10.20   & 4.26 \\  
    F356W  & 0.42 & 0.42 & 8.50    & 4.34 \\    
\hline    
\end{tabular}
\tablefoot{The background is given in Jy/px, since it is a constant for all non-zero pixels, amounting to at most $\sim$\,0.13\,Jy for F090W (and less for the other filters). The saturated pixels in F277W were filled with the peak flux density in the image.}
\end{table}

Integrated flux densities of images, galaxy models, constant background levels and residuals are reported in Table \ref{tab:fluxes}. Taken by themselves, the total fluxes of the masked jet in the residual images are broadly consistent with the radio-to-optical spectrum with a spectral index $\alpha = 0.7$--$1.0$, but amongst each other still follow a similar trend as the galaxy stellar bump, see Fig.~\ref{fig:SED}. Given that some pixels are saturated, the reported jet fluxes constitute lower bounds on the total emission. 

\begin{figure}
    \centering

    \includegraphics[width=\columnwidth]{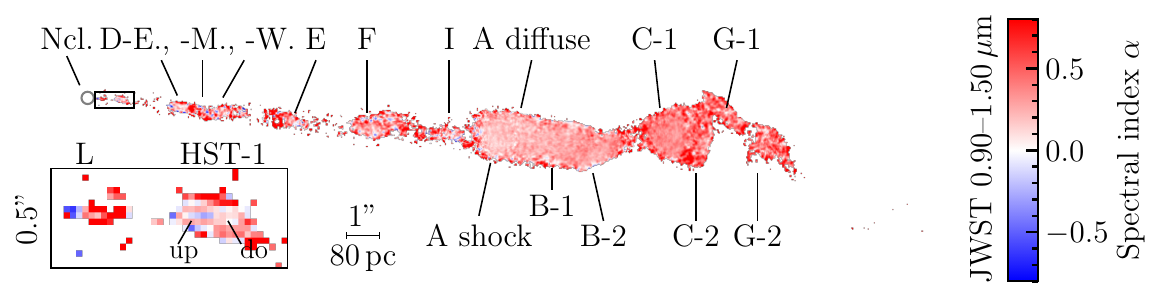}
    \includegraphics[width=\columnwidth]{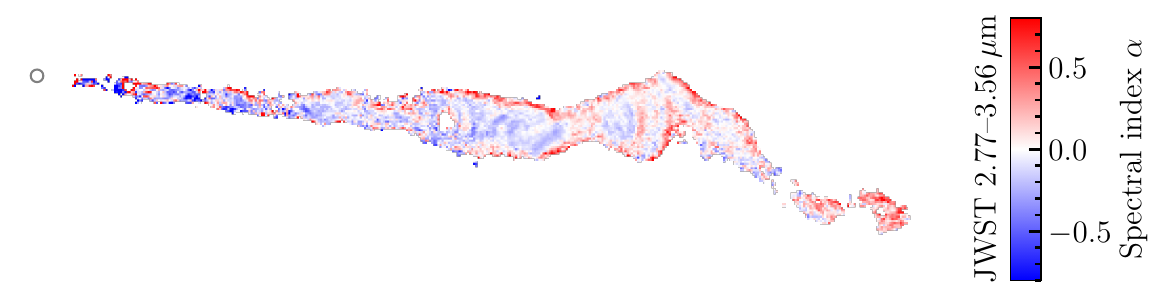}
    \caption{Spectral index maps of the isolated M87 infrared jet between the short and long wavelength images, respectively (top and bottom panels). The index $\alpha$ is defined as $S_\lambda \propto\lambda^\alpha\ (\propto \nu^{-\alpha})$. Floors of 0.5\% (top panel) and 1.5\% (bottom panel) of the image maxima were applied prior to calculating the spectral index. The images have the same spatial scale.}
    \label{fig:spectralindex}
\end{figure}

To obtain the resolved spectral index maps, the 0.9--1.5\,$\upmu$m and 2.77--3.56\,$\upmu$m image pairs were first aligned in Fourier space to correct for residual translation offsets between filters. Between 0.90\,$\upmu$m and 1.50\,$\upmu$m, the jet is generally optically thin ($\alpha > 0$; see top panel in Fig.~\ref{fig:spectralindex}); in most of the individual components leading up to the shock (component A), the spectral index is almost flat. Beyond A and B, the spectrum appears to steepen mildly, likely due to synchrotron cooling \citep[e.g.,][]{Snios2019ApJ...879....8S}.

Component L is characterized by a strong spectral index gradient, with $\alpha$ growing in the downstream direction. HST-1 shows a mean $\alpha = 0.05$, but the two sub-components are clearly distinct: the upstream one shows $\alpha_{\rm up} = -0.15$, and the downstream one $\alpha_{\rm do} = 0.30$. This further supports the multi-component structure of HST-1, as indicated in Fig.~\ref{fig:jet}. Indeed, this structure was studied already by \cite{Biretta1999}, and HST-1 was recently modelled with two components also in Chandra X-ray observations \citep{Thimmapa2024}. Both overall flattening and gradient of the spectral index within the knots support a recollimation shock nature of these features \citep{Laing2013}.

At the longer wavelengths, the spectral index is mostly flat along the jet, if not mildly negative. For parts of L, HST-1, D-East, and A, no spectral index can be obtained since the pixels are saturated in the 2.77\,$\upmu$m image. Similar to the short wavelengths, the spectrum steepens in the extended emission region, downstream from the C component. As an alternative way to visualise the spectral structure of M87 we provide the multi-filter image mapped onto red, green, blue (RGB) space in Appendix~\ref{sec:rgb}.

\section{Discussion and conclusion}

Since the synchrotron loss dominated jet sustains a power law spectrum across many orders of magnitude in wavelength and spatial scale, the radiative losses must be somewhat balanced by the internal particle acceleration mechanism. \cite{Lesch1998ApJ...499..167L} proposed continuous magnetic reconnection-driven particle acceleration to achieve the necessary energies for optical emission, concluding that in order for this process to be dominant, jets must contain magnetic flux filaments across appropriate scales. 

Indeed, jets are often characterized by a large-scale helical filamentary substructure \citep[e.g.,][]{Fuentes2023NatAs...7.1359F}; this is also the case for M87 on the scales probed by the observations presented in this work. The helical jet structure is not necessarily indicative of a similarly structured magnetic field, even though both are present in the M87 jet and coincide well \citep{Pasetto2021ApJ...923L...5P}. The magnetic field structure may be deduced from the polarised emission, which cannot be recovered from NIRCam observations. However, components D, E, F and I of our JWST images follow the known double helix in M87, shaped by Kelvin-Helmholtz instabilities \citep[KHI; e.g.,][]{Lobanov2003NewAR..47..629L,Hardee2011ApJ...735...61H,Nikonov2023MNRAS.526.5949N}, specifically their elliptical surface mode, consistent with \cite{Pasetto2021ApJ...923L...5P}.

Ultimately, jets decelerate through interactions with the interstellar medium and the conversion of bulk kinetic into internal energy, accelerating particles and dissipating energy as radiation \citep[e.g.,][]{Laing2014MNRAS.437.3405L}. In our M87 images, this occurs at the projected distance of $\sim$2\,kpc from the nucleus. \cite{Fichet2025A&A...693A.270F} showed that the distribution of gas in the galaxy may influence the extent of the radio emission region, whereas the stellar population can interact with the jet via stellar wind driven mass loading, influencing the dissipation region. 

In this work, we presented the first JWST+NIRCam images of the M87 jet at 0.90, 1.50, 2.77 and 3.56\,$\upmu$m. Our observations fall into the 1.6\,$\upmu$m infrared bump, dominated by stellar emission; while this bias could not be entirely removed during the galaxy subtraction process, the residual jet images are broadly consistent with the radio to optical spectrum. In the higher resolution shorter wavelength images, we identified all distinct jet components up to $\sim$\,24\,arcsec from the nucleus, including \mbox{HST-1}, the more upstream knot L, and the bright shock farther downstream. The images and spectral index maps provided further evidence for a complex substructure of HST-1, as indicated by previous works. In the 2.77 and 3.56\,$\upmu$m images we also detect the counter-jet $\sim$\,24\,arcsec from the nucleus and found its morphology to be consistent with radio observations.

\begin{acknowledgements}
We thank the TRGB-SBF Project team, as well as J. L. Gómez, T. Savolainen and S. Markoff for comments on the manuscript.
JR further acknowledges financial support from the Severo Ochoa grant CEX2021-001131-S funded by MCIN/AEI/10.13039/501100011033. MW is supported by a Ramón y Cajal grant RYC2023-042988-I from the Spanish Ministry of Science and Innovation. JJ and GSA acknowledge financial support from STScI grant JWST GO-3055. This work used observations made with the NASA/ESA/CSA James Webb Space Telescope. The data were obtained from the Mikulski Archive for Space Telescopes at the Space Telescope Science Institute, which is operated by the Association of Universities for Research in Astronomy, Inc., under NASA contract NAS 5-03127 for JWST. These observations are associated with JWST program GO-3055. 

This research has made use of the NASA/IPAC Extragalactic Database, which is funded by the National Aeronautics and Space Administration and operated by the California Institute of Technology (\url{https://doi.org/10.26132/NED1}).

This research has made use of the VizieR catalogue access tool, CDS,
Strasbourg, France \citep{10.26093/cds/vizier}. The original description 
of the VizieR service was published in \citet{vizier2000}.

\end{acknowledgements}

\bibliographystyle{aa}

\bibliography{bibliography}

\appendix 
\onecolumn

\section{Model subtraction and masking}
\label{sec:processing}

As detailed in Section \ref{sec:obs+data}, each image is processed by subtracting a constant background level measured far from the galaxy centre, along with the smooth galaxy model fitted in elliptical annuli. 
We additionally fitted a smooth model to the residual image after the galaxy model was subtracted to remove any residual galaxy structure and to ensure that the background level was exactly zero. Lastly, a mask is applied to spatially isolate the jet region from the residual emission caused by stellar surface brightness fluctuations. 

The calibration procedure, requiring very accurate characterisation of the JWST-NIRCam instrument and including modelling of the galactic component, may be subject to residual systematics that are difficult to identify. In particular, the inverted spectrum implied by the F277W and F356W measurements is a puzzling result that is difficult to reconcile with optically thin synchrotron emission, and would require a contribution from Compton scattering. Future improvements to the calibration of the instrument will clarify these issues.

\begin{figure*}[!htbp]
    \centering
    \includegraphics[width=0.82\textwidth]{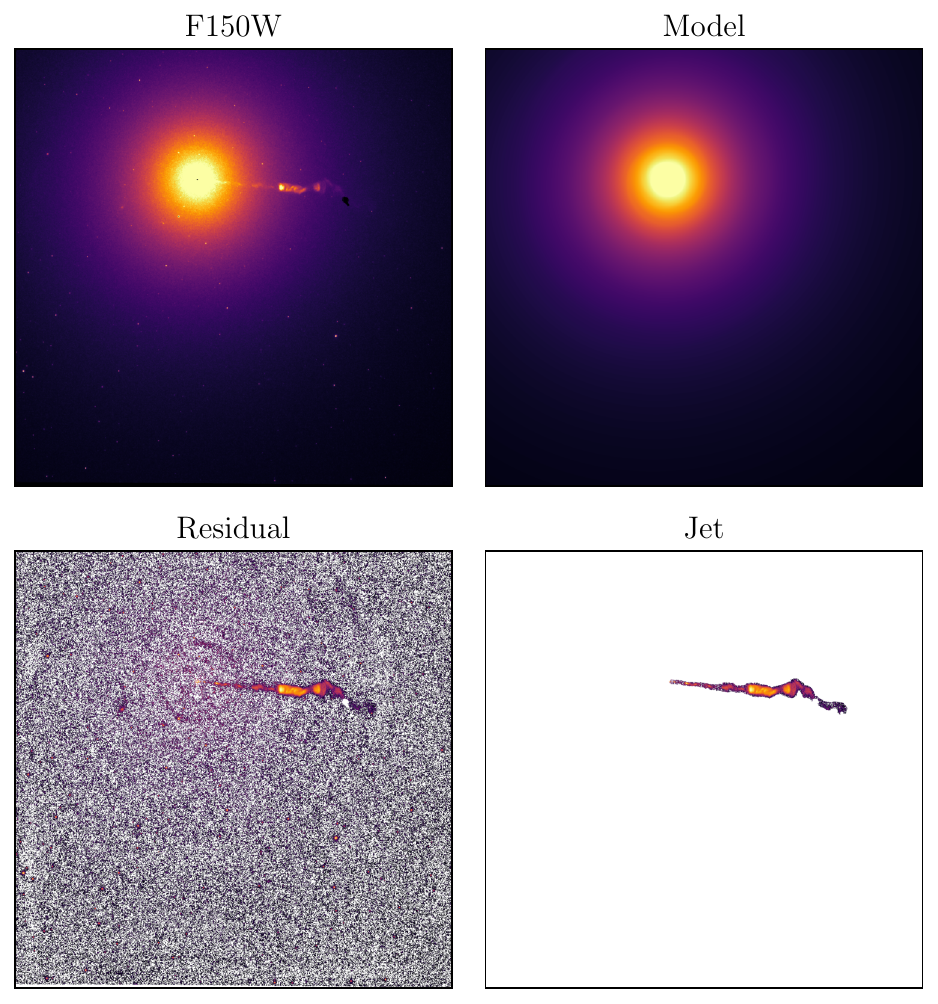}
    \caption{Processing of the F150W image: Full F150W image, smooth galaxy model, residual, and masked jet image. The prominent patch in the jet bend, appearing black in the full image and white in the residual, is a clump of bad pixels in the A1 detector. The two upper images are on a linear, the other two on a logarithmic colour scale; white pixels in the residual image indicate negative residual flux values following the subtraction of the mean component of the galaxy model.}
    \label{fig:overview}
\end{figure*}

\clearpage

\section{Counter-jet}
\label{sec:counterjet}

\begin{figure*}[!htbp]
    \centering
    \includegraphics[width=0.82\textwidth]{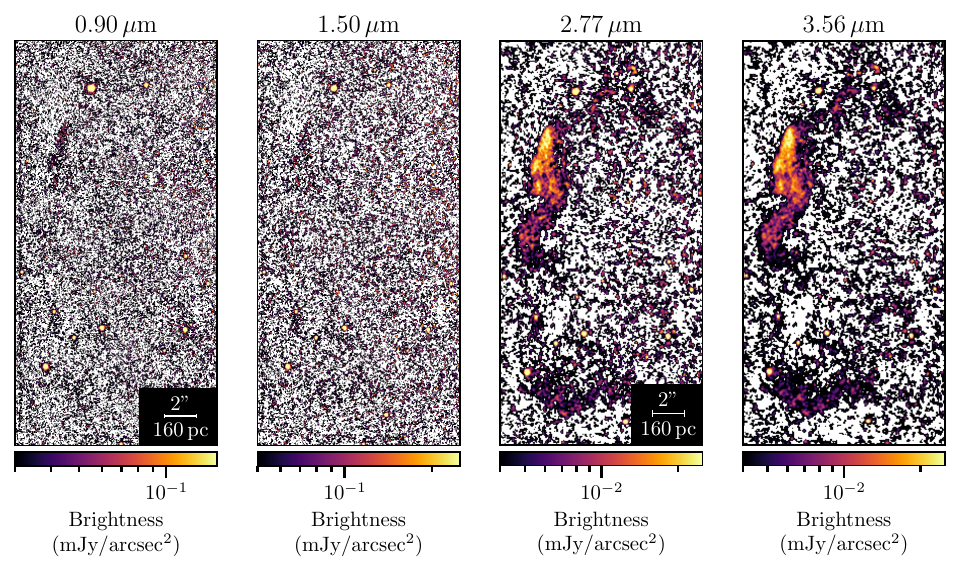}
    \caption{Counter-jet in the JWST images of M87. The images are shown in logarithmic colour scale, with limits chosen for best visibility of the counter-jet; the dynamic ranges are 5 for the short, and 6.25 for the long wavelengths. The negative pixels were coloured white instead of black to further increase the contrast.}
    \label{fig:counterjet}
\end{figure*}

\begin{figure*}[!htbp]
    \centering
    \includegraphics[width=0.82\textwidth]{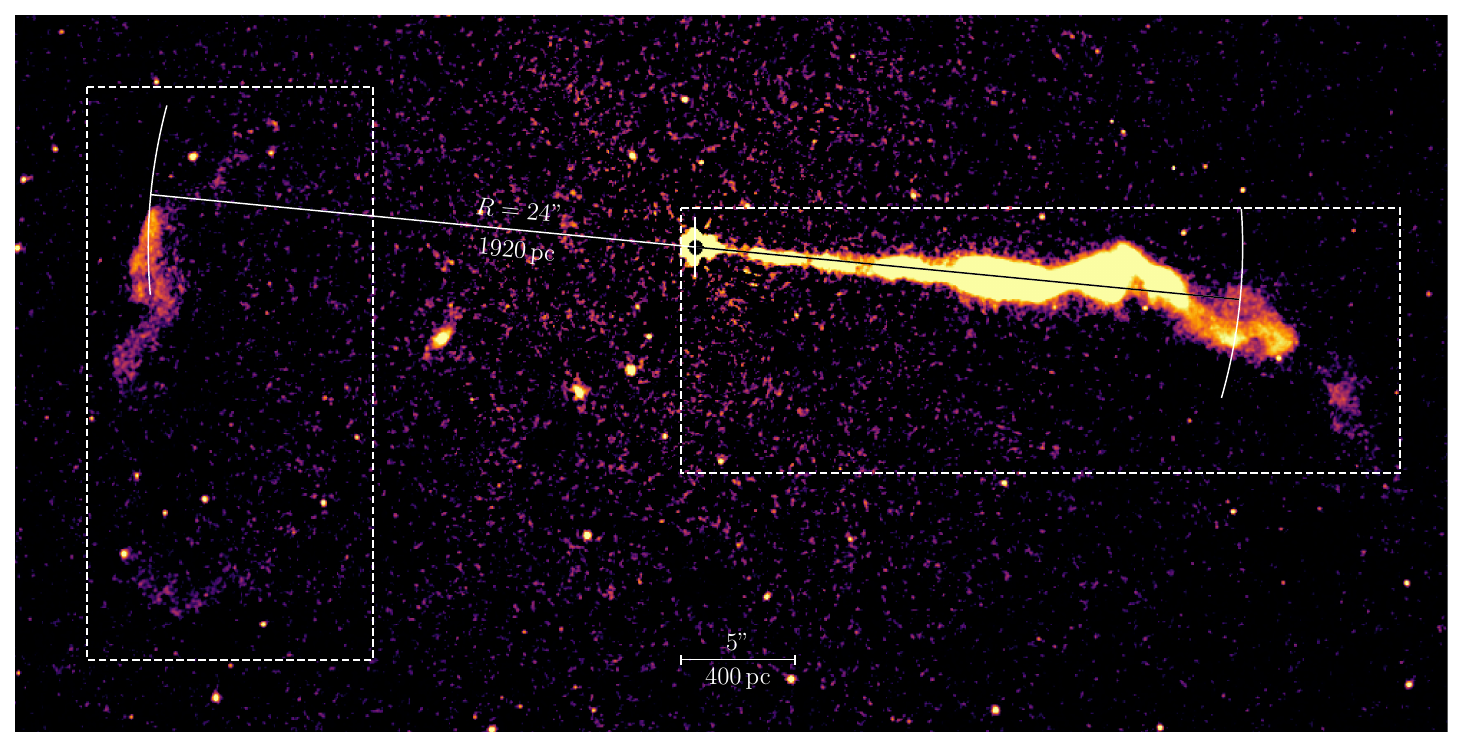}
\caption{Regions in the 3.56\,$\upmu$m residual image used to analyse the jet (Fig.~\ref{fig:jet}) and counter-jet (Fig.~\ref{fig:counterjet}). The hot spot component of the counter-jet appears about 24\,arcsec away from the nucleus.}
    \label{fig:boxes}
\end{figure*}

We detect faint emission from the C-shaped counter-jet in all wavelength bands, most prominently in the longer wavelength images, see Fig.~\ref{fig:counterjet}. The counter-jet consists of two filaments connected by a hot spot, which is the only (barely) visible part in the shorter wavelengths. Figure \ref{fig:boxes} indicates the image regions used for the presented analysis, as well as the relative position of the jet and counter-jet components. The hot spot component of the counter-jet appears 24\,arcsec away from the nucleus, transitioning into a corkscrew shape below (right panels in Fig.~\ref{fig:counterjet}). This morphology is consistent with VLA observations \citep{Pasetto2021ApJ...923L...5P}.

Along the forward jet, the faint extended emission further downstream from the jet knots labelled in Fig.~\ref{fig:jet} is located at the same 24\,arcsec distance from the nucleus, likely corresponding to deflected jet material, decelerating through interactions with the interstellar medium. The upstream part of the counter-jet is not visible, presumably due to strong Doppler de-boosting. We measured the total flux density of the observed counter-jet component to be ($230\pm 30$)\,$\upmu$Jy at 3.56\,$\upmu$m, and ($220\pm 30$)\,$\upmu$Jy at 2.77\,$\upmu$m.

\section{RGB image of M87}
\label{sec:rgb}

In order to simultaneously visualise observations recorded with multiple filters, we obtained an RGB image of M87 using the F356W, F150W and F090W data mapped onto the red, green and blue channels. Bad pixels were filled with the mean flux of good pixels in each colour layer. The image is aligned with the celestial equatorial coordinate system (north is up).

\begin{figure*}[!htbp]
    \centering
    \includegraphics[width=\textwidth]{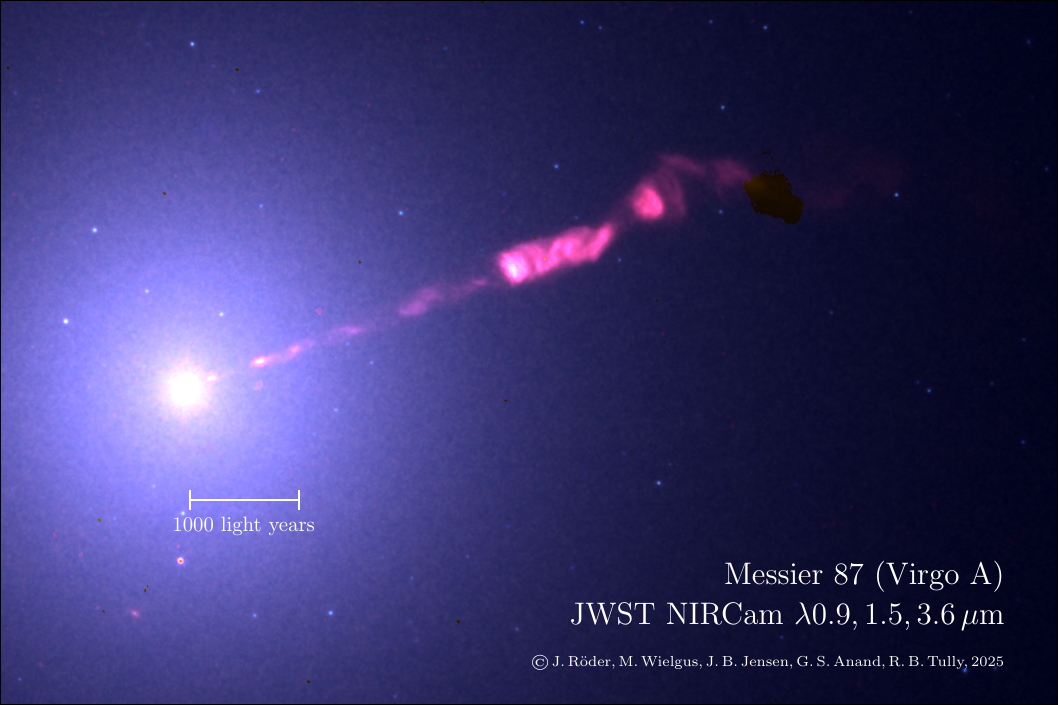}
    \caption{RGB image of M87 obtained using F356W, F150W and F090W observations.}
    \label{fig:rgb}
\end{figure*}

\clearpage

\end{document}